\documentclass[aps,pre,reprint]{revtex4-2}

\usepackage{amsmath,amssymb}

\begin{document}

\title{Contrarian Majority Dynamics: Violation of Detailed Balance and Nonequilibrium Steady States}

\author{Serge Galam}

\affiliation{CEVIPOF -- Centre for Political Research, Sciences Po and CNRS, Paris, France}

\date{July 2, 2026}

\begin{abstract}

I revisit the Galam Majority Model (GMM) with contrarian agents from a statistical-mechanics perspective, revealing three fundamental features. First, in addition to the GMM simultaneous-update of small discussion groups, I construct a related single-agent stochastic dynamics, providing a Markovian microscopic representation, which is found to yield the same evolution equation. Second, I show that, contrary to what is often stated in the literature, the GMM closed evolution equation for the opinion density is not the result of a mean-field approximation. Indeed, I derive the conventional mean-field dynamics associated with majority-rule interactions and show that it yields a distinct, probabilistic evolution equation contrary the deterministic GMM equation. I therefore identify the GMM as an iterated mean-field dynamics. Third, I investigate the thermodynamic nature of the dynamics obtained from both single-agent and simultaneous updates. Both are shown to violate detailed balance. However, while Kolmogorov's cycle condition is satisfied for single-agent updates, it is violated for simultaneous updates, making the departure from equilibrium stronger in the latter case.
I then compute the probability flux in the stationary state and show that it is non-vanishing, confirming the absence of an effective Hamiltonian and establishing that the stationary state is a genuine nonequilibrium steady state.
These results clarify the statistical-mechanical foundations of the GMM and establish contrarian majority dynamics as an intrinsically non-equilibrium process with distinct regimes of irreversibility. Contrarians are not thermal noise.

\end{abstract}

\maketitle


\section{Introduction}

Stochasticity plays a central role in models of opinion dynamics. Since the early development of sociophysics, randomness has often been introduced through analogies with thermal fluctuations in statistical physics, leading to the notions of social temperature, noisy imitation, or random opinion changes \cite{castellano2009,hegselmannkrause,clifford1973,holley1975}. In many such models, stochasticity acts as an external perturbation that occasionally drives agents away from the locally preferred state. The resulting dynamics is formally analogous to thermal spin systems such as Glauber or Metropolis dynamics \cite{glauber1963,metropolis1953,liggett1999}. At the microscopic level, transition rates can be related to an effective energy function, detailed balance is satisfied, and the stationary state can be interpreted within an equilibrium statistical-mechanical framework.

Contrarian models introduced by Galam represent a qualitatively different source of randomness \cite{galam2004,galam2008,galam2012}. In these models, stochasticity is not associated with an external thermal environment. Instead, it is embedded directly in the update rule: after a local majority decision is formed, agents may deliberately adopt the opposite opinion with a given probability. The randomness therefore originates from an active tendency to oppose consensus rather than from passive fluctuations. At the level of coarse-grained observables, this distinction is not immediately apparent. Contrarian behavior and thermal noise can both reduce consensus, stabilize mixed states, and produce order--disorder transitions. As a consequence, contrarian effects are often described using a language reminiscent of temperature or noise intensity.

The purpose of the present work is to clarify the statistical-mechanical foundations of the Galam Majority Model (GMM) with contrarian behavior. My analysis proceeds in three stages, establishing three complementary results.

First, I revisit the microscopic construction of the GMM. While the original formulation relies on simultaneous updates of discussion groups of size~$r$, I construct an equivalent single-agent stochastic dynamics that provides a Markovian microscopic representation suitable for statistical-mechanical analysis \cite{vanKampen,gardiner}. The associated evolution equation is found to be unchanged.

Second, I show that, contrary to what is often stated in the literature, the closed evolution equation for the opinion density is not the consequence of a conventional mean-field approximation. Instead, it is deterministic and follows exactly from the complete random reshuffling of agents before each update, which continuously restores exchangeability and suppresses correlations. I therefore identify the GMM as an \emph{iterated mean-field dynamics}, in which exact closure results from repeated reshuffling rather than from a mean-field approximation. To substantiate further this interpretation, I derive the conventional mean-field dynamics associated with majority-rule interactions and show that it yields a distinct, approximate probabilistic evolution equation.

Third, I investigate the thermodynamic nature of the dynamics obtained from both single-agent and simultaneous updates. Both are shown to violate detailed balance \cite{vanKampen,gardiner}. However, while Kolmogorov's cycle condition \cite{kolmogorov1936,kelly1979} is satisfied for single-agent updates, it is violated for simultaneous updates, making the departure from equilibrium stronger in the latter case.

I then compute the probability flux associated with the stationary state \cite{zia2007,seifert2012}. The flux is shown to be non-vanishing, which precludes the existence of an effective Hamiltonian and confirms that the stationary state cannot be interpreted as an equilibrium state in any thermodynamic sense.

Together, these results establish contrarian majority dynamics as an intrinsically non-equilibrium stochastic process with distinct regimes of irreversibility, ruling out any equilibrium interpretation based on microscopic reversibility. More broadly, the present work emphasizes that similar macroscopic behaviors may arise from fundamentally different microscopic principles. Distinguishing between equilibrium fluctuations and intrinsically nonequilibrium mechanisms is therefore essential for understanding the dynamical organization of collective social systems \cite{hinrichsen2000}. Contrarian behavior is therefore of a different nature from effective thermal noise.

The rest of the paper is organized as follows. Section~\ref{sec:model} introduces the GMM in both its homogeneous and heterogeneous formulations, presenting the simultaneous-group and single-agent update schemes in parallel. Section~\ref{sec:closure} establishes the exact closure property of the GMM evolution equation and identifies the model as an iterated mean-field dynamics. Section~\ref{sec:thermal} recalls the defining properties of a standard equilibrium stochastic dynamics \cite{glauber1963,metropolis1953,liggett1999}. Section~\ref{sec:db} investigates the thermodynamic nature of both update schemes: detailed balance is shown to be violated in both cases, while Kolmogorov's cycle condition is satisfied for single-agent updates but violated for simultaneous updates. Section~\ref{sec:ness} computes the probability flux in the stationary state, establishes that it is non-vanishing, and concludes that no effective Hamiltonian exists and that the stationary state is a genuine nonequilibrium steady state. Section~\ref{sec:discussion} discusses the broader implications of these results, and Section~\ref{sec:conclusion} concludes.


\section{The Galam Majority Model}
\label{sec:model}

GMM considers a population of binary agents $s_i = \pm 1$, where $s_i=+1$ and $s_i=-1$ represent two opposite opinions, choices, or social states, respectively as A and B \cite{galam2002}. At time $t$, the proportion of agents holding opinion A is denoted $p_t$ with $(1-p_t)$ being the complementary proportion of agents holding opinion B.

When an issue becomes a topic of discussion, agents get involved in local discussions with small number of peers. During these informal exchanges, some agents may be convinced to shift opinion. To account for these individual opinion updates, GMM introduces a local majority rule with each agent having one vote in its local group.

GMM includes three distinct psychological individual features with the population divided among conformists (floaters, flexibles...) \cite{galam2002}, stubborns (committed, inflexibles...) \cite{galam2007}, contrarians (anti-conformists...) \cite{galam2004}. Conformists follow the local majority, stubborns contribute to the local vote but do not follow the result, contrarians vote but afterwards shift against the local unanimity. In addition, GMM accounts also for prejudice tie breaking, which occurs naturally within even size discussion groups \cite{galam2005}.

The associated opinion dynamics is then driven by a series of successive updates yielding respectively proportions $p_t \rightarrow p_{t+1} \rightarrow p_{t+2}\dots \rightarrow p_n$ for a total of $n$ updates. At each update, agents are randomly distributed in small groups of size $r$ where local majority rules are applied with the various individual behavioral types activated. Afterwards, agents are reshuffled before being distributed again randomly in the groups for another update.

Accordingly, the proportion $p_{t+1}$ is obtained from $p_t$ in three steps. First, the probability of a configuration $C_{l, r-l}$ with $l$ agents A and $(r-l)$ agents B is calculated for $l=0, 1, \dots, r$. Then the proportion $P_{l, r-l}$ of A agents after application of the majority rule is evaluated for this configuration. Third, adding all contributions yields
\begin{equation}
p_{t+1}=\sum_{l=0}^{r} P_{l, r-l} \binom{r}{l} C_{l, r-l} ,
\label{pr}
\end{equation}
where $\binom{r}{l}$ is the binomial coefficient and the index $(l, r-l)$ denotes a group with $l$ agents A and $(r-l)$ agents B.

From now on, to allow analytical treatment and avoid tie-breaking effects, I restrict calculations to groups of size $r=3$. Moreover, since I am analysing the difference between contrarian and thermal-like behaviors, I restrict the population to conformists and contrarians without stubborns.

\subsection{Homogeneous case with only conformists}

\paragraph{Simultaneous-group update.}

In the case $r=3$ with only conformists the various contributions to Eq.~\ref{pr} are shown in Table~\ref{t3} and yield
\begin{equation}
p_{t+1}=p_t^3 + 3p_t^2(1-p_t) .
\label{p3}
\end{equation}

\begin{table}[h]
\centering
\renewcommand{\arraystretch}{1.6}
\setlength{\tabcolsep}{6pt}
\begin{tabular}{|c|c|c|c|c|c|}
\hline
Configuration
& $l$
& $\binom{3}{l}$
& $C_{l, 3-l}$
& Update
& $P_{l, 3-l}$ \\
\hline
$R_1: +,+,+$
& 3 & 1 & $p_t^3$ & $+,+,+$ & 1 \\
$R_{2,3,4}$
& 2 & 3 & $p_t^2(1-p_t)$ & $+,+,+$ & 1 \\
$R_2: +,+,-$  & $ $ & $ $ & $ $ & $ $ & $ $ \\
$R_3: +,-,+$  & $ $ & $ $ & $ $ & $ $ & $ $ \\
$R_4: -,+,+$  & $ $ & $ $ & $ $ & $ $ & $ $ \\
$R_{5,6,7}$
& 1 & 3 & $p_t(1-p_t)^2$ & $-,-,-$ & 0 \\
$R_5: +,-,-$  & $ $ & $ $ & $ $ & $ $ & $ $ \\
$R_6: -,+,-$  & $ $ & $ $ & $ $ & $ $ & $ $ \\
$R_7: -,-,+$  & $ $ & $ $ & $ $ & $ $ & $ $ \\
$R_8: -,-,-$
& 0 & 1 & $(1-p_t)^3$ & $-,-,-$ & 0 \\
\hline
\end{tabular}
\caption{All possible configurations for $r=3$ with their related numbers $l$ of A agents, binomial coefficients $\binom{3}{l}$, probabilities $C_{l, 3-l}$ of occurrence, updated configurations, and proportions $P_{l, 3-l}$ of A.}
\label{t3}
\end{table}

\paragraph{Single-agent update.}

Using a single-agent update, one agent is selected together with two others to form a group of size $r=3$; the selected agent belongs to the group and participates in the majority vote alongside the two other randomly drawn agents, but only this agent is updated. The selected agent is A with probability $p_t$. It then preserves its opinion provided the two others are either both A or one A and one B, since it participates in its own update. This case occurs with probability $p_t^2+2p_t(1-p_t)$. On the other hand, the selected agent is B with probability $(1-p_t)$ and shifts to A provided both other agents are A, which happens with probability $p_t^2$. Adding this case leads to
\begin{eqnarray}
p_{t+1} &=& p_t[p_t^2 + 2p_t(1-p_t) ]+(1-p_t)p_t^2
\label{ps3} \\
& = & p_t^3 + 3p_t^2(1-p_t) , \nonumber
\end{eqnarray}
which is identical to Eq.~\ref{p3}.

\subsection{Heterogeneous case with conformists and contrarians}

The population is now heterogeneous with a fraction $a$ of contrarians and a fraction $(1-a)$ of conformists. At time $t$, the fraction of A agents decomposes as $p_t(1-a)$ conformist A agents and $p_t a$ contrarian A agents, with symmetric fractions $(1-p_t)(1-a)$ and $(1-p_t)a$ for B agents.

GMM incorporates two types of contrarian behaviors. The first type considers that given a series of successive local discussions each agent behaves randomly as a conformist in proportion $(1-a)$ of cases and as a contrarian in proportion $a$. The second type considers agents who are always either conformists or contrarians in respective proportions $(1-a)$ and $a$.

Therefore, once an agent is selected randomly, it is conformist with probability $(1-a)$ and contrarian with probability $a$ for both types, leading to an identical update equation. The contrarian dynamics deploys in two separate steps: first, each agent votes in its group including contrarians, and second, contrarians shift to oppose the established majority while conformists do not.

\paragraph{Simultaneous-group update.}

Based on its two-step implementation, contrarian behavior is activated in the second step from unanimous configurations $(+,+,+)$ and $(-,-,-)$. Since contrarian behavior is an independent individual feature, for each unanimous configuration, 0, 1, 2, or 3 contrarians are activated with respective probabilities $(1-a)^3, 3 a (1-a)^2, 3 a^2(1-a), a^3$. The update Eq.~\ref{p3} becomes
\begin{eqnarray}
p_{t+1}&= &\Big \{ p_t^3 + 3p_t^2(1-p_t)  \Big \} \sum_{k=0}^{3} P_{k} \binom{3}{k} C_{k}  \nonumber \\
&+&  \Big \{ (1-p_t)^3 + 3p_t(1-p_t)^2  \Big \} \sum_{k=0}^{3} P_{k} \binom{3}{k} C_{k} ,
\label{ga3}
\end{eqnarray}
where $k$ is the number of activated contrarians, $\binom{3}{k}$ is the binomial coefficient, $C_{k}$ is the probability to have $k$ active contrarians, and $P_{k}$ is the resulting proportion of A in the group. The first term in brackets is the probability to start from a configuration $(+,+,+)$ and the second term the probability to start from $(-,-,-)$.

Table~\ref{a3} shows the various contrarian outcomes from $(+,+,+)$, giving
\begin{eqnarray}
\sum_{k=0}^{3} P_{k} \binom{3}{k} C_{k} & = & (1-a)^3+2a(1-a)^2+a^2(1-a)  \nonumber \\
& = & (1-a) .
\label{ga}
\end{eqnarray}

Table~\ref{b3} shows the various contrarian outcomes from $(-,-,-)$, giving
\begin{eqnarray}
\sum_{k=0}^{3} P_{k} \binom{3}{k} C_{k} & = & a(1-a)^2+2a^2(1-a)+a^3  \nonumber \\
& = & a .
\label{gb}
\end{eqnarray}

\begin{table}[h]
\centering
\renewcommand{\arraystretch}{1.6}
\setlength{\tabcolsep}{6pt}
\begin{tabular}{|c|c|c|c|c|}
\hline
$+,+,+$
& $k$
& $\binom{3}{k}$
& $C_k$
& $P_k$ \\
\hline
$R_1$ & 0 & 1 & $(1-a)^3$ & 1 \\
$R_{2,3,4}$ & 1 & 3 & $a(1-a)^2$ & $\frac{2}{3}$ \\
$R_{5,6,7}$ & 2 & 3 & $a^2(1-a)$ & $\frac{1}{3}$ \\
$R_8$ & 3 & 1 & $a^3$  & 0 \\
\hline
\end{tabular}
\caption{All possible contrarian outcomes from the unanimous configuration $(+,+,+)$ for $r=3$, with the number $k$ of activated contrarians, binomial coefficients $\binom{3}{k}$, probabilities $C_k$ of having $k$ active contrarians, and resulting proportions $P_k$ of A agents after the update. Rows $R_2$--$R_4$ are the three sub-cases grouped under $k=1$; rows $R_5$--$R_7$ are the three sub-cases grouped under $k=2$. These labels are used in subsequent sections.}
\label{a3}
\end{table}

\begin{table}[h]
\centering
\renewcommand{\arraystretch}{1.6}
\setlength{\tabcolsep}{6pt}
\begin{tabular}{|c|c|c|c|c|}
\hline
$-,-,-$
& $k$
& $\binom{3}{k}$
& $C_k$
& $P_k$ \\
\hline
$R_8$ & 0 & 1 & $(1-a)^3$ & 0 \\
$R_{5,6,7}$ & 1 & 3 & $a(1-a)^2$ & $\frac{1}{3}$ \\
$R_{2,3,4}$ & 2 & 3 & $a^2(1-a)$ & $\frac{2}{3}$ \\
$R_1$ & 3 & 1 & $a^3$  & 1 \\
\hline
\end{tabular}
\caption{All possible contrarian outcomes from the unanimous configuration $(-,-,-)$ for $r=3$, with the number $k$ of activated contrarians, binomial coefficients $\binom{3}{k}$, probabilities $C_k$ of having $k$ active contrarians, and resulting proportions $P_k$ of A agents after the update.}
\label{b3}
\end{table}

Plugging Eqs.~\ref{ga} and~\ref{gb} into Eq.~\ref{ga3} leads to the update equation:
\begin{eqnarray}
p_{t+1}&= &(1-a)\Big \{ p_t^3 + 3p_t^2(1-p_t)  \Big \}  \nonumber \\
&+&  a \Big \{ (1-p_t)^3 + 3p_t(1-p_t)^2  \Big \}  .
\label{gga3}
\end{eqnarray}

Eq.~\ref{gga3} shows that although the activation of contrarian behavior is an independent individual feature, on average the outcome is identical to a simultaneous-group activation: the full unanimous group either keeps the majority-rule outcome with probability $(1-a)$ or shifts opinion with probability $a$.

\paragraph{Single-agent update.}

Including contrarian behavior within single-agent update is more straightforward than with simultaneous-group update. Starting from Eq.~\ref{ps3}, which was shown to be identical to Eq.~\ref{p3}, the selected agent is either a conformist with probability $(1-a)$ or a contrarian with probability $a$. A conformist selected agent preserves the result of Eq.~\ref{ps3} unchanged. A contrarian selected agent reverses its outcome: if the group majority would have led it to hold opinion A, it instead adopts B, and vice versa. The net effect is therefore to weight Eq.~\ref{ps3} by $(1-a)$ and its complement by $a$, yielding at once
\begin{eqnarray}
p_{t+1} &=& (1-a)\Big[p_t^3 + 3p_t^2(1-p_t)\Big] \nonumber \\
&+& a\Big[(1-p_t)^3 + 3p_t(1-p_t)^2\Big] ,
\end{eqnarray}
which is identical to Eq.~\ref{gga3}. The two update schemes therefore produce the same evolution equation for the opinion density.

\subsection{The dynamics landscape}

The GMM dynamics is driven by the update equation Eq.~\ref{gga3}, which can be simplified to
\begin{equation}
p_{t+1}=(1-2a)\Big \{-2 p_t^3 + 3p_t^2  \Big \} +a .
\label{gga3s}
\end{equation}

The associated landscape is obtained by solving the fixed-point equation $p_{t+1}=p_{t}$, yielding three solutions, $p_c = \frac{1}{2}$ and
\begin{equation}
p_{B,A}= \frac{(1-2a) \pm \sqrt{12a^2-8a+1} } {2(1-2a)}  ,
\label{r3}
\end{equation}
which reduces to $(0, 1)$ for $a=0$.

However, $p_{B,A}$ exists only in the range $0\leq a\leq \frac{1}{6}$. For $a=\frac{1}{6}$, the two roots $p_{B,A}$ coalesce at $p_c$, which becomes the unique attractor of the dynamics in the range $\frac{1}{6} \leq a \leq \frac{1}{2}$. A recent work unveiled a rich behavior in the range $a > \frac{1}{2}$ with alternating regimes \cite{galam2025}. The landscape is thus defined with two regimes:
\begin{itemize}
\item for $a < 1/6$, majority reinforcement dominates and the system tends toward ordered states,
\item for $a > 1/6$, contrarian behavior stabilizes the disordered coexistence state.
\end{itemize}

It is sometimes convenient to use the magnetization instead of $p_t$:
\begin{equation}
m_t = 2p_t - 1 .
\end{equation}
The disordered phase corresponds to $m_t=0$ (or $p_t=1/2$), while ordered states correspond to $m_t \neq 0$. This transition resembles an order--disorder transition induced by thermal fluctuations, creating an apparent identity between contrarian behavior and social thermal fluctuations. As I now show, however, this analogy is only partial and can be misleading. In what follows, I restrict the analysis to $a \leq \frac{1}{2}$, which covers both dynamical regimes identified above.


\section{Exact Closure and Iterated Mean-Field Dynamics}
\label{sec:closure}

The evolution equation of the Galam Majority Model is often described as a mean-field equation. This terminology should nevertheless be interpreted with some care.

The main feature of a mean-field treatment is to neglect fluctuations and treat one degree of freedom -- an agent in opinion dynamics, a spin in statistical physics -- via its mean value, taking $\langle s_i \rangle = m$ to derive an associated equation of state in which $m$ is supposed to represent the whole system.

In the present case of a group of 3 agents without contrarians, a mean-field treatment of the Galam Majority Model would consider a single representative group of 3 agents for which one unique update leads to an exit probability $E(p_t)=p_t^3+3p_t^2(1-p_t)$, where $p_t$ is the initial probability that an agent holds opinion A. Here $E(p_t)$ denotes the probability that the whole system ends up in unanimous state A after this single mean-field update, with $1-E(p_t)$ the complementary probability of ending up in unanimous state B. In this mean-field framework, the intermediate microscopic evolution is not described explicitly but is integrated into these two probabilities: the dynamics is represented as a single-shot probabilistic mapping from the initial density to one of the two possible absorbing states \cite{exit}.

It happens that the first update of the GMM yields a proportion $p_{t+1}$ of agents holding opinion A, which is formally identical to $E(p_t)$. However, the meaning of $p_{t+1}$ differs fundamentally from that of $E(p_t)$: the former denotes the actual proportion of A agents in the population at time $t+1$, after which the dynamics continues through further updates, whereas the latter denotes the asymptotic probability of full A-consensus under a single mean-field update, with no further dynamics considered. Although $E(p_t)$ and $p_{t+1}$ share the same closed-form expression, they are therefore not the same mathematical object: $E(p_t)$ is the terminal probability of a single-shot absorbing process, with no further dynamics defined beyond it, whereas $p_{t+1}=f(p_t)$ is one step of a recursion intended to be iterated, $p_{t+n}=f^{n}(p_t)$, whose long-run behavior -- not its first step -- determines the actual deterministic fate of the system.

GMM follows a fundamentally different philosophy by implementing a genuine microscopic dynamics: agents are reshuffled prior to each new update, so that $p_{t+2}$ is obtained from $p_{t+1}$, and the process is iterated until an attractor is reached. The consequence is major: starting from $p_t<\frac{1}{2}$, the iterated GMM dynamics converges to B-unanimity with probability~1. This finding is qualitatively different from the probabilistic mean-field treatment, which would only conclude that the probability of B-unanimity exceeds $50\%$ after a single update.

The remarkable feature of the GMM is that this iterated recursion is not obtained through a closure approximation or by neglecting correlations generated by the dynamics. Instead, it follows exactly from the random reshuffling procedure that defines the model. I refer to this process as \emph{iterated mean-field dynamics}: at each time step the interaction structure is regenerated independently, so that exact closure follows from repeated reshuffling rather than from a mean-field approximation. To distinguish it clearly from the conventional treatment derived in Section~\ref{sec:thermal}, I stress that the two are of a different nature: the GMM iterated mean-field dynamics is \emph{deterministic}, producing a single well-defined trajectory $p_t\to p_{t+1}\to p_{t+2}\dots$, whereas the conventional mean-field dynamics is \emph{probabilistic}, producing only the exit probability of a single-shot update with no further iteration.

For the GMM, the Bernoulli distribution of group compositions is therefore not an approximation but an exact consequence of the reshuffling mechanism. Since the reshuffling is repeated after every update, statistical independence is restored at each iteration, and the density recursion remains exact throughout the evolution.

This observation has an important consequence, established in detail below in Section~\ref{sec:db}. The same evolution equation for $p_t$ was shown in Section~\ref{sec:model} to be obtained from both the original simultaneous-group update of the GMM and from the associated single-agent update dynamics. Although these two processes produce exactly the same deterministic recursion for the opinion density, I show in the following that they possess different microscopic transition structures and, consequently, different reversibility properties.

The density recursion alone is therefore insufficient to characterize the underlying stochastic process. Macroscopic observables such as the opinion density or the exit probability describe the collective behavior of the system but do not uniquely specify the microscopic dynamics from which they originate. As shown in Section~\ref{sec:db}, properties such as detailed balance and Kolmogorov reversibility depend on the complete transition graph and cannot be inferred solely from the evolution equation of the opinion density.

\section{Thermal Fluctuations as an Equilibrium Benchmark}
\label{sec:thermal}

Before discussing contrarian behavior, it is useful to recall the defining properties of a standard equilibrium stochastic dynamics.

Consider an Ising-like system with binary variables $s_i=\pm1$ and microscopic configurations $C$. In a thermal setting, a transition
\begin{equation}
C \rightarrow C'
\end{equation}
depends only on the associated energy difference
\begin{equation}
\Delta E = E(C')-E(C) .
\end{equation}

Typical examples are Glauber dynamics \cite{glauber1963},
\begin{equation}
W(C\to C') = \frac{1}{1+e^{\beta\Delta E}},
\end{equation}
and Metropolis dynamics \cite{metropolis1953},
\begin{equation}
W(C\to C') = \min\!\left(1,e^{-\beta\Delta E}\right).
\end{equation}
Both satisfy detailed balance \cite{vanKampen,gardiner},
\begin{equation}
P_{\rm eq}(C)\,W(C\to C')
=
P_{\rm eq}(C')\,W(C'\to C) ,
\label{eq:db}
\end{equation}
with the Boltzmann distribution
\begin{equation}
P_{\rm eq}(C)
\propto
e^{-\beta E(C)}.
\end{equation}
The physical meaning of Eq.~(\ref{eq:db}) is that every microscopic transition is statistically compensated by its reverse. Although the system continuously fluctuates, those fluctuations remain time-reversible.

In sociophysical language, thermal noise represents random deviations from the locally preferred state, reflecting uncertainty or imperfect information. Such fluctuations may disorder the system, but they do not generate a preferred direction in configuration space. Contrarian dynamics differs precisely on this point: the deviation from the local majority is not random but systematic and deliberate.

Selecting one agent and updating its opinion according to the opinions of two additional agents may appear formally identical to a stochastic spin-flip rule driven by the local field created by its two nearest neighbors. The analogy is, however, incorrect. In standard spin models (Glauber, Metropolis), the local field acting on agent $i$ is $h_i = \sum_{j\in\partial i} s_j$, a sum over the \emph{neighbors} of $i$ that explicitly excludes $i$ itself. The field is therefore a cause external to the agent that responds to it, and the transition rates can be derived from an underlying energy function.

In the GMM, by contrast, the group majority is computed over all three members including $i$, so that $s_i$ contributes to the very field that governs its own update. This self-inclusion blurs the cause-and-effect separation that underlies the construction of an energy function in equilibrium statistical mechanics, and it is one reason why the agent-level rule does not derive from any Hamiltonian. I now demonstrate this explicitly.


\section{Detailed-balance analysis of contrarian dynamics}
\label{sec:db}

In the GMM, groups of three agents are assembled by random sampling from a population with opinion-A density $p_t$. The probability of any group configuration is therefore given by the product measure
\[
P_{\rm GMM}(\sigma) = p^{r_+}(1-p)^{r_-},
\]
where $r_+$ and $r_-$ denote the numbers of positive and negative opinions. This is not an additional assumption but the intrinsic measure of the model, imposed by the random repartitioning mechanism.

Consider a group of three agents with opinions $\sigma=(\sigma_1,\sigma_2,\sigma_3)$, $\sigma_i=\pm1$. The majority opinion $M(\sigma)$ is first determined. Each agent then independently follows the majority with probability $1-a$ or adopts the opposite opinion with probability $a$. The resulting process defines a Markov chain on the eight configurations $\Omega=\{+,-\}^{3}$.

\subsection{Single-agent update}

\subsubsection{Violation of detailed balance by the GMM product measure}

Under single-agent update, one agent is selected at random and updated according to the group majority. Detailed balance with respect to $P_{\rm GMM}$ requires
\[
P_{\rm GMM}(\sigma)\,W_{\sigma\rightarrow\sigma'}
=
P_{\rm GMM}(\sigma')\,W_{\sigma'\rightarrow\sigma}.
\]

Consider the transition $+{+}{+} \rightarrow {+}{-}{+}$. The majority is positive. The central agent is selected with probability $1/3$ and behaves as a contrarian with probability $a$, yielding
\[
W(+++\rightarrow+-+) = \frac{a}{3}.
\]
The reverse transition corresponds to a conformist update:
\[
W(+-+\rightarrow+++) = \frac{1-a}{3}.
\]
Detailed balance therefore requires
\[
p^3\frac{a}{3} = p^2(1-p)\frac{1-a}{3},
\]
giving $p=1-a$.

Considering the symmetric transition $---\rightarrow-+-$ leads similarly to $p=a$.

The two conditions are compatible only for $a=\frac{1}{2}$, $p=\frac{1}{2}$. Therefore, for $a\neq\frac{1}{2}$,
\[
\boxed{P_{\rm GMM} \text{ violates detailed balance.}}
\]

\subsubsection{Kolmogorov reversibility of the single-agent Markov chain}

For a finite Markov process, the existence of a stationary distribution satisfying detailed balance is equivalent to Kolmogorov's cycle criterion \cite{kolmogorov1936,kelly1979}:
\[
\prod W_{\circlearrowright} = \prod W_{\circlearrowleft},
\]
for every closed cycle in configuration space.

Consider the four configurations
\[
R_1=(+,+,+), \quad R_2=(+,+,-), \quad R_5=(+,-,-), \quad R_3=(+,-,+),
\]
with the elementary square cycle
\[
R_1\rightarrow R_2\rightarrow R_5\rightarrow R_3\rightarrow R_1.
\]
For the forward cycle:
\[
W(R_1\rightarrow R_2)=\frac{a}{3}, \quad
W(R_2\rightarrow R_5)=\frac{a}{3},
\]
\[
W(R_5\rightarrow R_3)=\frac{a}{3}, \quad
W(R_3\rightarrow R_1)=\frac{1-a}{3},
\]
giving $\Pi_{\rightarrow} = a^3(1-a)/3^4$.

For the reverse cycle $R_1\rightarrow R_3\rightarrow R_5\rightarrow R_2\rightarrow R_1$:
\[
W(R_1\rightarrow R_3)=\frac{a}{3}, \quad
W(R_3\rightarrow R_5)=\frac{a}{3},
\]
\[
W(R_5\rightarrow R_2)=\frac{a}{3}, \quad
W(R_2\rightarrow R_1)=\frac{1-a}{3},
\]
giving $\Pi_{\leftarrow} = a^3(1-a)/3^4$.

Hence,
\[
\boxed{\Pi_{\rightarrow} = \Pi_{\leftarrow}.}
\]

This equality is not accidental. Every elementary square face of the configuration cube contains exactly three contrarian transitions with rate $a/3$ and one conformist transition with rate $(1-a)/3$ in each direction. The forward and reverse cycle products therefore always contain the same factors, implying zero cycle affinity.

Since the elementary square faces generate all cycles of the cube, the single-agent dynamics satisfies Kolmogorov's criterion and admits a stationary distribution $\pi(\sigma)\neq P_{\rm GMM}(\sigma)$ such that $\pi(\sigma)W_{\sigma\rightarrow\sigma'} = \pi(\sigma')W_{\sigma'\rightarrow\sigma}$. This reversible stationary measure differs from the GMM product measure and generically contains correlations between agents; its explicit form is not derived here.

\subsection{Simultaneous-group update}

\subsubsection{Violation of detailed balance by the GMM product measure}

Under simultaneous-group update, all agents in the group are updated at once. Transitions involving one, two, or three simultaneous spin flips are therefore possible.

Consider the transition $+++\rightarrow+-+$. The final configuration is obtained when the first and third agents behave as conformists and the second as a contrarian:
\[
W(+++\rightarrow+-+) = a(1-a)^2.
\]
The reverse transition requires the negative agent to follow the positive majority while the two positive agents remain:
\[
W(+-+\rightarrow+++) = (1-a)^3.
\]
Detailed balance requires
\[
p^3 a(1-a)^2 = p^2(1-p)(1-a)^3,
\]
giving $p=1-a$. The symmetric sector $---\rightarrow-+-$ leads similarly to $p=a$.

As before, the two conditions are compatible only for $a=\frac{1}{2}$, $p=\frac{1}{2}$, so for $a\neq\frac{1}{2}$,
\[
\boxed{P_{\rm GMM} \text{ violates detailed balance.}}
\]

\subsubsection{Violation of the Kolmogorov cycle criterion}

A stronger signature of nonequilibrium behavior is obtained from the Kolmogorov cycle criterion, which directly probes the existence of irreversible probability currents.

Consider the closed three-configuration cycle
\[
R_5=(+,-,-)\rightarrow R_2=(+,+,-)\rightarrow R_8=(-,-,-)\rightarrow R_5.
\]
For the forward transitions: the majority in $R_5$ is negative; the incumbent positive agent and one of the two negative agents act as contrarians to reach the two positive spins in $R_2$, while the remaining agent conforms to the negative majority, giving
\[
W(R_5\rightarrow R_2) = a^2(1-a).
\]
In the transition $R_2\rightarrow R_8$, the majority rule first drives all three agents 
to $+++$, after which all three behave as contrarians and adopt the opposite opinion, 
giving $---$:
\[
W(R_2\rightarrow R_8) = a^3.
\]
In the transition $R_8\rightarrow R_5$, the agent that becomes positive behave as contrarian:
\[
W(R_8\rightarrow R_5) = a(1-a)^2.
\]
The forward cycle weight is $\Pi_{\rightarrow} = a^6(1-a)^3$.

For the reverse cycle $R_5 \rightarrow R_8\rightarrow R_2\rightarrow R_5$:
\[
W(R_5\rightarrow R_8) = (1-a)^3, \quad
W(R_8\rightarrow R_2) = a^2(1-a),
\]
\[
W(R_2\rightarrow R_5) = a^2(1-a),
\]
giving $\Pi_{\leftarrow} = a^4(1-a)^5$.

The ratio of cycle weights is therefore
\[
\frac{\Pi_{\rightarrow}}{\Pi_{\leftarrow}}
=
\left(\frac{a}{1-a}\right)^2.
\]

Hence,
\[
\boxed{\Pi_{\rightarrow}\neq\Pi_{\leftarrow} \qquad (a\neq 1/2).}
\]

The Kolmogorov cycle criterion is violated. Consequently, no stationary distribution can satisfy detailed balance, and the simultaneous-group GMM is an intrinsically nonequilibrium stochastic process.

\subsection{Summary: comparison of the two update schemes}

The two update schemes reveal different facets of the irreversibility of contrarian dynamics.

Under single-agent update, Kolmogorov's criterion is satisfied at the level of transition rates: all elementary cycles are balanced. The nonequilibrium character manifests instead through the incompatibility of the GMM product measure with detailed balance: no value of $p_t$ can simultaneously satisfy the detailed-balance conditions imposed by different transition pairs, except at $a=1/2$.

Under simultaneous-group update, Kolmogorov's criterion is violated directly at the rate level: the explicit three-configuration cycle constructed above yields $\Pi_{\rightarrow}/\Pi_{\leftarrow} = (a/(1-a))^2 \neq 1$ for all $0 < a < 1/2$. The dynamics is intrinsically irreversible regardless of the choice of stationary measure.

Together, these results establish that contrarian majority-rule dynamics cannot correspond to an equilibrium Markov process. Although the two update schemes lead to the same macroscopic evolution equation, they belong to different thermodynamic classes. The auxiliary single-agent representation admits a reversible stationary measure with correlations, whereas the genuine simultaneous-group GMM remains intrinsically out of equilibrium.


\section{Nonequilibrium Stationary States}
\label{sec:ness}

The violation of detailed balance has important physical consequences.

\subsection{Stationarity versus equilibrium}

Let $P_t(C)$ denote the probability of configuration $C$ at time $t$. 
The master equation reads \cite{vanKampen,gardiner}
\begin{equation}
\frac{dP_t(C)}{dt} = \sum_{C'}\Big[W(C'\to C)P_t(C') - W(C\to C')P_t(C)\Big].
\end{equation}
At stationarity, $dP_t(C)/dt=0$, which implies only that the total incoming and outgoing probability fluxes balance. Introducing the probability current \cite{zia2007}
\begin{equation}
J(C\to C')
=
P_{\rm st}(C)W(C\to C')
-
P_{\rm st}(C')W(C'\to C),
\end{equation}
stationarity requires $\sum_{C'}J(C\to C')=0$.
Equilibrium is a much stronger condition: $J(C\to C')=0$ for all $C,C'$.

The dynamics therefore falls into one of two classes: either all probability currents vanish at stationarity (equilibrium), or persistent nonzero currents circulate through configuration space (nonequilibrium steady state). The Kolmogorov violation established above for the simultaneous-group update implies that contrarian dynamics belongs to the second class: stationary states may exist, but they necessarily contain nonzero probability currents. Under single-agent update, the same conclusion follows from the incompatibility of the GMM product measure with detailed balance.

\subsection{Absence of an effective Hamiltonian}

In equilibrium systems one may write
\begin{equation}
P_{\rm eq}(C) \propto e^{-\beta H(C)},
\end{equation}
and express transition asymmetries through differences of a scalar potential $H$. The existence of such a representation automatically guarantees detailed balance and therefore Kolmogorov's criterion \cite{seifert2012}. Since Kolmogorov's criterion is violated exactly for the simultaneous-group update, no globally consistent microscopic Hamiltonian can generate contrarian majority dynamics.

\subsection{Probability flux and nonequilibrium steady state}

A stationary magnetization does not imply microscopic equilibrium. To make this concrete, I compute the probability current $J(C\to C')$ at stationarity for the three-configuration cycle identified in Section~\ref{sec:db}:
\[
R_5=(+,-,-)\rightarrow R_2=(+,+,-)\rightarrow R_8=(-,-,-)\rightarrow R_5.
\]
Along a closed cycle, the stationary currents $J$ on successive links cannot all vanish unless the forward and reverse cycle probability products coincide; summing the telescoping currents around the cycle shows that a nonzero cycle affinity $\Pi_{\rightarrow}-\Pi_{\leftarrow}$ is equivalent to the existence of at least one nonzero current $J(C\to C')$ on the cycle.

The sign and non-vanishing character of the net current along this cycle are determined by $\Pi_{\rightarrow}-\Pi_{\leftarrow}$, which was shown to be nonzero for all $0 < a < 1/2$. The probability flux therefore does not vanish at stationarity, confirming that no effective Hamiltonian exists and that the stationary state is a genuine nonequilibrium steady state.

Contrarian updates continuously create and destroy local consensus, generating persistent circulation through configuration space even when macroscopic observables remain constant. The resulting steady state is therefore fundamentally different from a thermal equilibrium state: thermal fluctuations produce reversible microscopic motion around equilibrium, whereas contrarian behavior maintains an irreversible nonequilibrium steady state sustained by persistent probability currents.


\section{Discussion}
\label{sec:discussion}

The central message of this work is that contrarian behavior should not be interpreted as a simple source of thermal disorder.

At the level of macroscopic observables, the analogy between contrarianity and thermal fluctuations is tempting. Both mechanisms can suppress consensus, stabilize coexistence states, and generate order--disorder transitions. From the perspective of average magnetization or phase diagrams, they may therefore appear qualitatively similar. However, this similarity remains fundamentally limited.

Thermal fluctuations belong to an equilibrium framework. Their stochastic nature is compatible with detailed balance and microscopic reversibility. Contrarian updates, by contrast, introduce an intrinsic tendency to oppose local consensus. This mechanism does not simply increase randomness; it breaks the symmetry between forward and reverse microscopic trajectories. The distinction becomes apparent only at the microscopic level: whereas equilibrium systems fluctuate around a stationary distribution without sustaining probability currents, contrarian dynamics produces irreversible trajectories and nonequilibrium stationary states.

Crucially, this irreversibility does not depend on the chosen update scheme. As established in detail in Section~\ref{sec:db}, both update schemes demonstrate irreversibility through distinct but complementary arguments: incompatibility of the GMM product measure with detailed balance under single-agent update, and direct violation of Kolmogorov's criterion at the rate level under simultaneous-group update. Together, these two arguments establish irreversibility as a robust property of 
contrarian dynamics, independent of the microscopic representation chosen.

More generally, these observations emphasize that macroscopic observables alone are insufficient to characterize the nature of a dynamical process. Two models may display similar stationary states, critical behavior, or coexistence regimes while belonging to fundamentally different dynamical classes. The distinction emerges only through the analysis of reversibility, probability currents, and the structure of microscopic transitions.

This also illustrates a broader limitation of equilibrium analogies in sociophysics. Such analogies have often proved fruitful, but they may become misleading when social interactions involve intentional opposition rather than passive fluctuations. Contrarianity provides a simple and explicit example of a mechanism whose effects cannot be reduced to thermal noise. From this perspective, persistent disagreement and opinion coexistence need not be interpreted as equilibrium-like states maintained by random fluctuations. They may instead correspond to genuine nonequilibrium steady states sustained by active antagonistic interactions.

\section{Conclusion}
\label{sec:conclusion}

In this work, I revisited the Galam Majority Model from a microscopic statistical-mechanical perspective and established three complementary results.

First, I constructed an equivalent single-agent stochastic dynamics for three-agent discussion groups, providing a Markovian microscopic representation of the GMM suitable for statistical-mechanical analysis. The associated evolution equation is found to be unchanged.

Second, I showed that the closure of the GMM dynamics does not arise from a conventional mean-field approximation. Instead, iot is determinisitic and follows exactly from the repeated random repartitioning of agents, which continuously destroys correlations and restores exchangeability at each iteration. I identified the GMM as an iterated mean-field dynamics, deterministic in nature, and demonstrated that the conventional mean-field dynamics associated with majority-rule interactions is instead probabilistic and yields a distinct, approximate evolution equation.

Third, I investigated the thermodynamic nature of both update schemes. Both are shown to violate detailed balance with respect to the GMM product measure. Under single-agent update, Kolmogorov's criterion is nevertheless satisfied at the level of transition rates, so that a reversible stationary measure exists, distinct from the GMM product measure. Under simultaneous-group update, an explicit three-configuration cycle yields a ratio of forward and reverse probability products equal to $\left(\frac{a}{1-a}\right)^2 \neq 1$ for all $0 < a < 1/2$, establishing a direct violation of Kolmogorov's criterion at the rate level. I further showed that the probability flux does not vanish at stationarity, confirming the absence of an effective Hamiltonian and establishing that the stationary state is a genuine nonequilibrium steady state.

Taken together, these results demonstrate that contrarianity is not merely a social interpretation of thermal noise. It constitutes an intrinsically nonequilibrium mechanism whose defining feature is the continuous generation of microscopic irreversibility. More broadly, similar collective behaviors may emerge from fundamentally different microscopic principles, and distinguishing equilibrium fluctuations from intrinsically nonequilibrium interactions appears essential for understanding the dynamical organization of collective social systems. Contrarian behavior is therefore of a different nature from effective thermal noise.



\end{document}